\documentclass[11pt, a4paper, logo, copyright]{googledeepmind}

\pdftrailerid{redacted}

\makeatletter
\renewcommand\bibentry[1]{\nocite{#1}{\frenchspacing\@nameuse{BR@r@#1\@extra@b@citeb}}}
\makeatother

\usepackage{amsmath,amsfonts,bm}

\def\eqref#1{equation~\ref{#1}}

\def\1{\bm{1}}

\DeclareMathAlphabet{\mathsfit}{\encodingdefault}{\sfdefault}{m}{sl}
\SetMathAlphabet{\mathsfit}{bold}{\encodingdefault}{\sfdefault}{bx}{n}

\newcommand{\softmax}{\mathrm{softmax}}

\usepackage{kantlipsum, lipsum}
\usepackage{dsfont}
\usepackage{dm-colors}

\usepackage{adjustbox}
\usepackage{todonotes}

\usepackage[authoryear, sort&compress, round]{natbib}
\usepackage[linesnumbered,ruled,vlined]{algorithm2e}
\SetCommentSty{mycommfont}

\graphicspath{{figures/}}

\title{Stealing User Prompts from Mixture of Experts}

\correspondingauthor{itayona@google.com}

\paperurl{}

\renewcommand{\today}{}

\usepackage{xspace}

\usepackage[frozencache=true,cachedir=myminted]{minted}
\usepackage{wrapfig}
\usepackage{array}  %
\usepackage{cellspace}  %
\usepackage{placeins}

\usepackage[linesnumbered,ruled,vlined]{algorithm2e}

\SetCommentSty{mycommfont}
\graphicspath{{figures/}}
\usepackage{xspace}
\newcommand{\attackname}{{MoE Tiebreak Leakage\xspace}}

\newcommand{\ecr}{{Expert-Choice-Routing\xspace}}
\newcommand{\moe}{{Mixture-of-Experts\xspace}}

\usepackage{subcaption}
\usepackage{hyperref}
\usepackage{url}
\usepackage{booktabs}

\usepackage{xcolor} %
\usepackage{colortbl} %
\usepackage{svg}
\usepackage{comment} %
\usepackage{cleveref}

\DeclareFixedFont{\ttb}{T1}{txtt}{bx}{n}{12} %
\DeclareFixedFont{\ttm}{T1}{txtt}{m}{n}{12}  %
\usepackage{xspace}

\newcommand{\topk}{{\texttt{topk}\xspace}}
\newcommand{\batchsize}{$B$}
\newcommand{\numexperts}{$N$}
\newcommand{\numlayers}{$D$}
\newcommand{\capacityfactor}{$\gamma$}
\newcommand{\expertcapacity}{$K$}
\newcommand{\seqlen}{$L$}
\newcommand{\seclen}{$M$}
\newcommand{\padlen}{$P$}
\newcommand{\totaltokens}{$\text{\batchsize{}} \times \text{\seqlen{}}$}
\newcommand{\vocab}{$\mathcal{V}$}
\newcommand{\vocabsize}{$V$}
\newcommand{\bvocab}{$\beta$}

\newcommand{\bseqT}{$\Phi$}

\newcommand{\onlinecomplexity}{$\mathcal{O}(\text{\vocabsize{}} \text{\seclen{}}^2)$}
\newcommand{\offlinecomplexity}{$\mathcal{O}(2^{\text{\numlayers} \text{\numexperts}} \text{\vocabsize{}} \text{\seclen{}} ^ 2)$}

\newcommand{\probeinput}{probe sequence}
\newcommand{\longsequence}{padding sequence}
\newcommand{\blockingsequence}{blocking sequence}

\newcommand{\advbatch}{\textit{{\color{magenta}adv\_batch}\xspace}}
\newcommand{\usersec}{\textit{{\color{olive}victim\_message}\xspace}}
\newcommand{\offlinemodel}{\textit{{\color{violet}local\_model}\xspace}}
\newcommand{\onlinemodel}{\textit{{\color{red}target\_model}\xspace}}
\usepackage{listings}

\author[1]{Itay Yona}
\author[1]{Ilia Shumailov}
\author[1]{Jamie Hayes}
\author[1]{Nicholas Carlini}

\affil[1]{Google DeepMind}

\begin{abstract}
\moe{} (MoE) models improve the efficiency and scalability of dense language models by \emph{routing} each token to a small number of \emph{experts} in each layer. In this paper, we show how an adversary that can arrange for their queries to appear in the same batch of examples as a victim's queries can exploit \ecr{} to fully disclose a victim's prompt. We successfully demonstrate the effectiveness of this attack on a two-layer Mixtral model, exploiting the tie-handling behavior of the \texttt{torch.topk} CUDA implementation. Our results show that we can extract the entire prompt using \onlinecomplexity{} queries (with vocabulary size \vocabsize{} and prompt length \seclen{}) or 100 queries on average per token in the setting we consider. This is the first attack to exploit architectural flaws for the purpose of extracting user prompts, introducing a new class of LLM vulnerabilities.
\end{abstract}

\begin{document}

\maketitle

\begin{figure*}[!htb]
\centering
\includegraphics[width=0.8\textwidth, trim={0 0 80 0},clip]{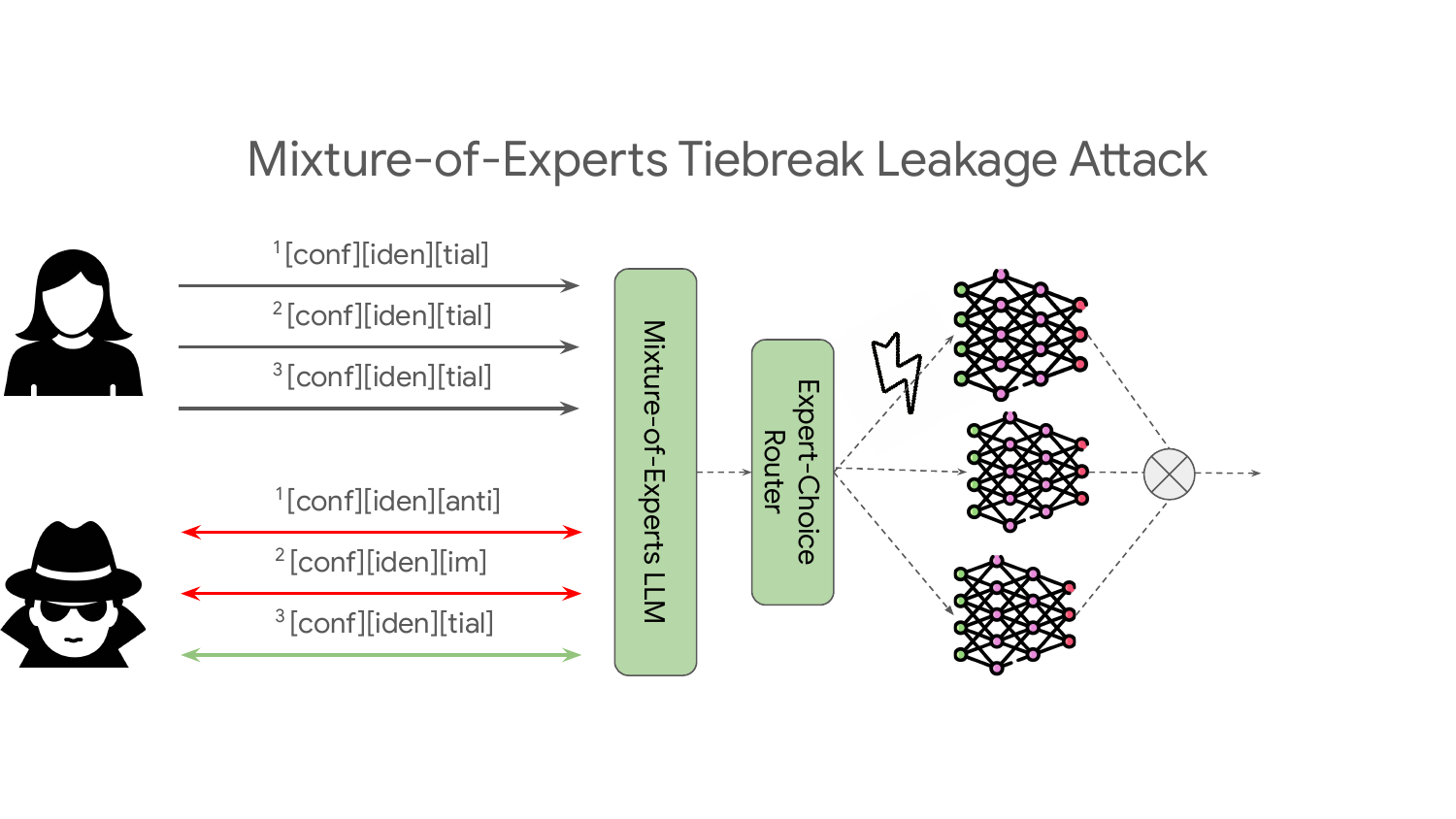}
\caption{High-level outline of the \attackname{} Attack. The attacker and victim queries are batched together, affecting routing of each other. The attacker systematically guesses the next token in a victim's confidential message. A correct guess triggers \ecr{} tie-breaker, leaving a detectable signal in the model's output.}
\label{fig:very_highlevel_attack_desc}
\end{figure*}

\section{Introduction}
Mixture-of-Experts (MoE) architectures have become increasingly important for large language models (LLMs) to handle growing computational demands \citep{shazeer2017outrageously,fedus2022switch,riquelme2021scaling,du2022glam}. This approach distributes the processing workload at each layer across multiple ``expert'' modules, allowing the model to selectively activate only the necessary experts for a given input. This selective activation improves efficiency and enables the development of larger, more capable LLMs. However, this approach can also introduce new vulnerabilities.

\cite{hayes2024buffer} recently identified a vulnerability in MoE models resulting from ``token dropping", which refers to a phenomenon that occurs when an expert's capacity is exceeded, causing excess tokens to be discarded or rerouted. They demonstrated that an adversary can exploit this vulnerability by strategically placing their data within the same batch as a victim's data, effectively overflowing the target expert buffers relied upon by the victim. This targeted overflow degrades the quality of the victim's model responses, resulting in a Denial-of-Service (DoS) attack.

\begin{figure*}[!htb]
\centering
\includegraphics[width=\textwidth]{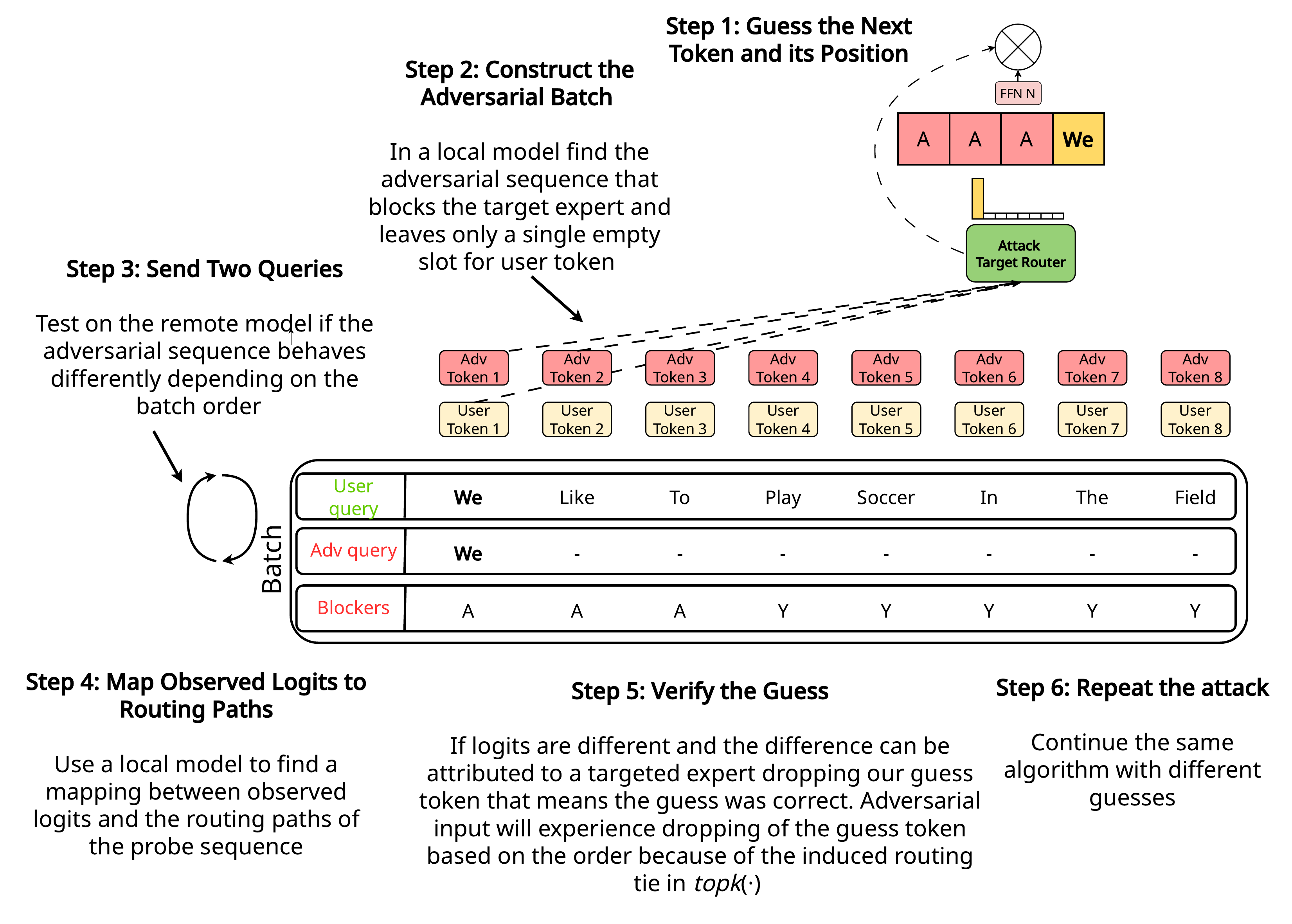}
\caption{Step-by-step execution of the \attackname{} attack. More details are provided in~\Cref{ssec:attackoverview}. }
\label{fig:highlevel_attack_desc}
\end{figure*}

In this paper we expand on this vulnerability, and demonstrate a novel attack with even more severe consequences: the complete disclosure of a victim's private input. By strategically crafting a batch of inputs, an attacker can manipulate the expert routing within a MoE model to leak the victim's prompt as we demonstrate in~\Cref{fig:very_highlevel_attack_desc}. This attack exploits the cross-batch side-channel created by token-dropping, where one user's data can influence the processing of another's. This influence, while subtle, creates an information leak that can be exploited to reveal the victim's input. Specifically, we show that MoE models using Expert Choice Routing (ECR) \citep{zhou2024expertchoicerouter} are vulnerable to this attack, dubbed \attackname{}. While our analysis focuses on ECR, the underlying principle of exploiting cross-batch dependencies suggests that other MoE routing mechanisms may be similarly vulnerable. We show the execution of the attack visually in~\Cref{fig:highlevel_attack_desc}.

This paper makes the following core contributions:

\begin{itemize}
    \item We introduce \attackname{} attack, a novel attack that exposes the vulnerability of user prompts in MoE models and highlights the critical need to consider security implications during architectural design.
    \item We demonstrate the feasibility of this attack on a Mixtral model \citep{jiang2024mixtralexperts} employing the ECR strategy. Our attack can verify a guessed user prompt with just two queries to the target model. The general attack requires \onlinecomplexity{}  queries to the target model and \offlinecomplexity{} queries to a local copy of the model, scaling with the number of experts (\numexperts{}), layers (\numlayers{}), vocabulary size (\vocabsize{}), and the length of the input sequence (\seclen{}).
    \item We discuss potential defense strategies to mitigate this vulnerability and enhance the security of MoE models.
\end{itemize}

\section{Background}

\subsection{Primer on Language Models and \moe{}}\label{ssec:primer}
A Transformer-based Large Language Model (LLM) is a function $f_{\theta}:\mathcal{V}^{\text{\seqlen}}\rightarrow \mathcal{P}(\mathcal{V})$ that takes as input a sequence of \emph{tokens} from a vocabulary $\mathcal{V}$ and outputs a probability distribution over the vocabulary, $\mathcal{P}(\mathcal{V})$. 
In particular, we are interested in functions of the form $f_{\theta}(z)=\softmax(W\cdot h_{\theta}(z))$, where $W$ is an unembedding matrix and $W\cdot h_{\theta}(z)$ gives a set of \emph{logits} over $\mathcal{V}$.

We assume that the model $h_{\theta}$ consists of multiple MoE layers.
An MoE layer consists of \numexperts{} expert functions $\{e_1, e_2, \ldots, e_{\text{\numexperts}}\}$ where $e_i: \mathbb{R}^d\rightarrow\mathbb{R}^d$ is a feed forward layer that takes in $d$-dimensional token representations and outputs new features of the same dimensionality. 
The MoE layer also consists of a gating function $g: \mathbb{R}^d\rightarrow\mathbb{R}^{\text{\numexperts}}$ which is used to assign token representations to experts by outputting a probability distribution over the \numexperts{} experts.

LLMs commonly process \emph{batches of inputs} to improve hardware utilization and efficiency.
This means $f_{\theta}$ in reality operates on the domain $\mathcal{V}^{{\text{\batchsize}}\times {\text{\seqlen}}}$, where \batchsize{} is the batch size and \seqlen{} is the sequence length of an input. 
For models that do not use MoE layers, the computation is entirely parallel over a batch of inputs; the computations of one input in the batch cannot affect the computations of another input in the batch.
For models that do use MoE layers, this is no longer true, as the gating function $g$ can only assign a limited number of token representations from a batch to a specific expert.
There are many different choices for how to assign tokens to experts given the output of the gating function; this is often called the \emph{routing strategy} \citep{cai2024surveymixtureexperts}. 

In this work, we focus on \ecr{}, which allows each expert to independently select its \topk{} assigned tokens from a batch of tokens \citep{zhou2024expertchoicerouter}.
The value \expertcapacity{} represents the fixed capacity of each expert, signifying the number of tokens it can process;
we refer to this as the \emph{expert's buffer capacity}.
This inherently ensures a balanced load across experts and introduces flexibility in allocating computational resources.
In our experimental setup, we define \expertcapacity{} as:
\begin{equation}\label{eq: exp_cap}
    \text{\expertcapacity} = \frac{\text{\totaltokens} \times \text{\capacityfactor}}{\text{\numexperts}}.
\end{equation}

Here, \totaltokens{} represents the total number of tokens in the input batch, $\text{\capacityfactor}>0$ is the capacity factor, indicating the average number of experts each token utilizes, and $\numexperts{}$ is the total number of experts.
Let $Z \in \mathbb{R}^{\text{\totaltokens} \times d}$ denote a batch of input token representations at a given layer, where $d$ is the hidden dimension of the model. 
For each $z_i\in Z$, we compute $g(z_i) = \{p_{i1}, p_{i2}, \ldots, p_{i\text{\numexperts}}\}$, which outputs a probability distribution over the \numexperts{} experts.
This produces the matrix:
\begin{align}
    G = \begin{bmatrix}
p_{1,1}       &   p_{1,2}       & \dots     &   p_{1,\text{\numexperts}}       \\
p_{2,1}       &   p_{2,2}       & \dots     &   p_{2,\text{\numexperts}}       \\
\vdots  &  \vdots   &   \vdots  &   \vdots  \\   
p_{\text{\totaltokens},1}       &   p_{\text{\totaltokens},2}       & \dots     &   p_{\text{\totaltokens},\text{\numexperts}}       \\
            \end{bmatrix},
\end{align}

where $p_{ij}$ represents the probability of assigning token $z_i$ to expert $e_j$.
\ecr{} applies a column-wise top-\expertcapacity{} selection of tokens; token $z_i$ is routed to expert $e_j$ if $p_{ij}$ is one of the top-\expertcapacity{} probabilities in column $j$.
Unlike other routing strategies, where experts may handle a variable number of tokens~\citep{fedus2022switch, lepikhin2020gshard, shazeer2017outrageously}, in \ecr{} the expert load is perfectly balanced by design, with each expert handling exactly \expertcapacity{} tokens.

Observe that not all tokens within the batch may be processed by an expert.
For example, if \capacityfactor{} is small (e.g. $<<1$) then the number of tokens processed by each expert is substantially smaller than \totaltokens{}, the total number of tokens in the batch.
In such cases, tokens that are not assigned to any expert are dropped -- that is, not processed by any expert~\citep{fedus2022switch, hwang2023tutel}.
This is commonly assumed to be of little consequence, as it is standard for MoE models to have residual connections between layers, meaning that the effect of dropping a token is limited.
However, we will show that token-dropping introduces a shared information side channel which can be exploited.

\subsection{Threat Model}\label{ssec:threat}
Having described the mechanism of token dropping in MoE models, we now define the threat model for our attack. This model, though simplified, represents a critical first step in understanding a new class of vulnerabilities in MoE systems. By establishing how token dropping can be exploited to leak user information, we aim to encourage future research into other potential attacks arising from design choices optimized for efficiency.

We make the following three simplifying assumptions.
\textbf{First}, we assume that the adversary has white-box access to the model that uses an MoE with cross-batch \ecr{} strategy~\citep{zhou2024expertchoicerouter}. This can apply in a setting where a third party is using the base model that is available publicly, e.g. implementation is available through t5x \citep{roberts2022t5x}. 

\textbf{Second}, the adversary can control the placement of its and the user inputs in the batch.
\textbf{Third}, the adversary can query the model repeatedly, ensuring that the user-supplied input is consistently in the same batch as its own inputs;
the adversary and user inputs are always batched together and sent to the model for processing. 

While the current attack requires strong assumptions, future work could explore techniques to relax these requirements. This might involve investigating methods to influence batch composition through timing attacks or through features in the model's serving infrastructure. We defer a more detailed discussion of the practicalities of the attack and potential methodological improvements to \Cref{sec:discussion}.

\section{\attackname{} Attack}
\label{sec:methodology}
We now proceed to describe the information leakage vulnerability (Section \ref{ssec:infoleakvuln}), the attack primitives (Section \ref{ssec:prims}), and two variants of our attack:

\begin{itemize}
\item \textbf{Oracle Attack:} Confirms whether a candidate prompt matches the victim's using only two queries.

\item \textbf{Leakage Attack:} Extracts the victim's prompt without any prior knowledge by iteratively applying the oracle attack to deduce the prompt token by token.
\end{itemize}

\subsection{Information-Leakage Vulnerability}
\label{ssec:infoleakvuln}
The vulnerability arises when a target token we aim to extract falls precisely at the boundary of an expert's capacity. By strategically submitting a ``guess'' token, we can influence the model's routing decisions to reveal whether our guess is correct.

\textbf{Incorrect guess:} The model's routing remains consistent for both our guess and the target, irrespective of their order in the input batch.

\textbf{Correct guess:} The model now perceives them as equivalent. Due to the expert's capacity constraint, their order in the batch becomes the deciding factor for processing. This behavioral discrepancy manifests as an observable difference in the model's output.

\newcommand{\drop}{\textbf{\textcolor{red}{Drops}}}
\newcommand{\notdrop}{\textbf{\textcolor{orange}{Doesn't drop}}}

\begin{table}[h!]
\centering
\begin{tabular}{|c|c|c|}
\hline
\textbf{Prioritization} & \textbf{Order in the Batch} & \textbf{Target token} \\ \hline
\bm{$ P_{guess} > P_{target} $} & First & \notdrop \\ \hline
\bm{$P_{guess} > P_{target} $} & Second & \notdrop \\ \hline
\bm{$P_{guess} < P_{target} $} & First & \drop \\ \hline
\bm{$P_{guess} < P_{target} $} & Second & \drop \\ \hline
\rowcolor{gray!20} \bm{$ P_{guess} = P_{target} $} & First & \notdrop \\ \hline
\rowcolor{gray!20} \bm{$P_{guess} = P_{target} $} & Second & \drop \\ \hline
\end{tabular}

\caption{Demonstrating how the token-dropping depends on both the relative priority of our guess compared to the target, and their order within the input batch. When their priority is equal, the input order acts as a tie-breaker, creating the information leak (See \Cref{sec:topkdeterminsm} for essential implementation details).}
\label{table:tiehandling}
\end{table}

We can exploit this by observing whether our guess is dropped by the expert. This allows us to iteratively probe for the hidden target token by submitting guesses and analyzing the model's response.

In essence, the information leakage vulnerability stems from ECR's decision process, which prioritizes tokens based on their identity but resolves ties using their order. An attacker can exploit this position-dependence by manipulating the input batch and observing whether the model's output changes. This allows the attacker to infer if their guess is equivalent to other tokens in the batch (i.e., if a tie occurred), effectively leaking information about the victim's prompt.

\subsection{Attack Primitives}
\label{ssec:prims}
To carry out the attack, we rely on three key primitives:

\begin{itemize}
\item \textbf{Controlling Expert Capacity:} The adversary extends the expert buffer capacity by including a \emph{padding sequence} - a long, arbitrary sequence in the adversarial batch. This ensures that the victim's tokens are not dropped by default and enables predictable tie-breaking behavior.
\item \textbf{Controlling Target Token Placement:} The adversary uses pre-computed \emph{blocking sequences} - sequences of tokens with high affinity for a specific expert - to fill the expert's buffer up to a desired position, leaving a single spot for the target token. Further details in \Cref{sec:findingblockers}.
\item \textbf{Recovering Target Token Routing Path:} In models with multiple MoE layers, the adversary needs to recover the \emph{routing path} of the target token to accurately interpret the effects of token dropping. This involves estimating the path of the known prefix using a local copy of the model and employing a ``routing-paths model'' to map different logits to their corresponding routing paths. Further details in \Cref{sec:recoverroutingpath}.
\end{itemize}

Using the first two primitives, we craft a dedicated adversarial batch to achieve the conditions described in Section \ref{ssec:infoleakvuln}. We illustrate the structure of this batch in Figure \ref{fig:adversarial_batch} and visualize its effect on the expert buffers in Figure \ref{fig:exploited_expert_buffers}.

\definecolor{mygold}{HTML}{B8860B}
\definecolor{mygreen}{HTML}{30A045}
\definecolor{myblue}{HTML}{296FBF}
\definecolor{myred}{HTML}{DF3E39}
\definecolor{mybrown}{HTML}{82645A}

\begin{figure}[!htb]
  \begin{center}
    \includegraphics[width=0.6\linewidth]{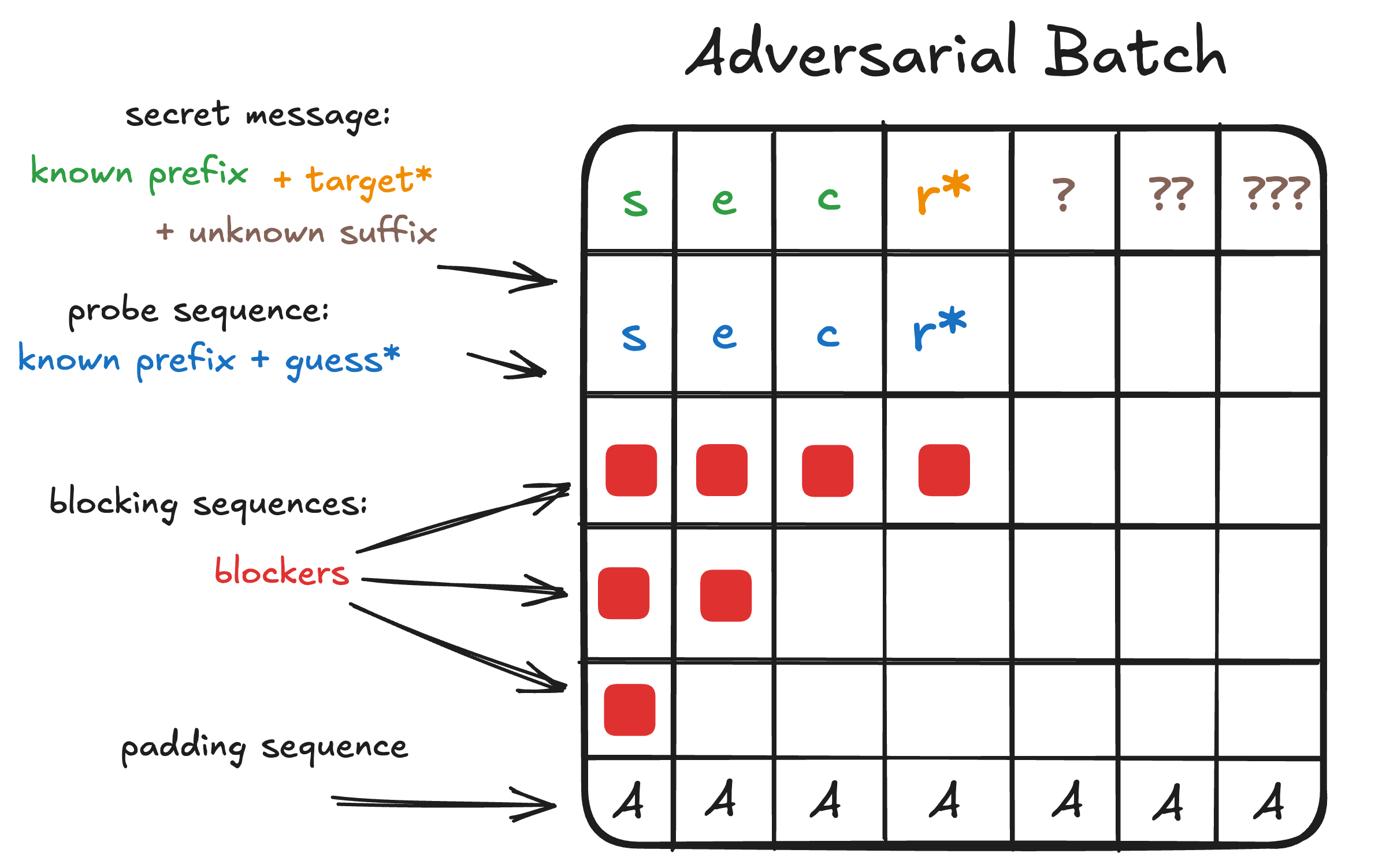}
  \end{center}

\caption{The adversarial batch consists of four components:
(1) \textbf{secret message} that attacker tries to leak, contains an already \textcolor{mygreen}{known prefix} a \textcolor{mygold}{target token} and an \textcolor{mybrown}{unknown suffix}.
(2) \textcolor{myblue}{\textbf{probe input}}, an attacker controlled sequence in which the known prefix and a guess for the target token are being sent. It aims to induce ties in \ecr{}, and for its returned output to be examined by the attacker for verification of correct guesses.
(3) \textcolor{myred}{\textbf{blocking sequences}}, a set of attacker controller inputs that aim to deprioritize the target and guess token, such that they will be placed at the edge of an expert buffer.
(4) \textbf{padding sequence}, an attacker controlled arbitrarly long sequence aims to extend the expert capacity (expert buffer length).}
\label{fig:adversarial_batch}
\end{figure}

\begin{figure}[!htb]
  \begin{center}
    \includegraphics[width=0.6\linewidth]{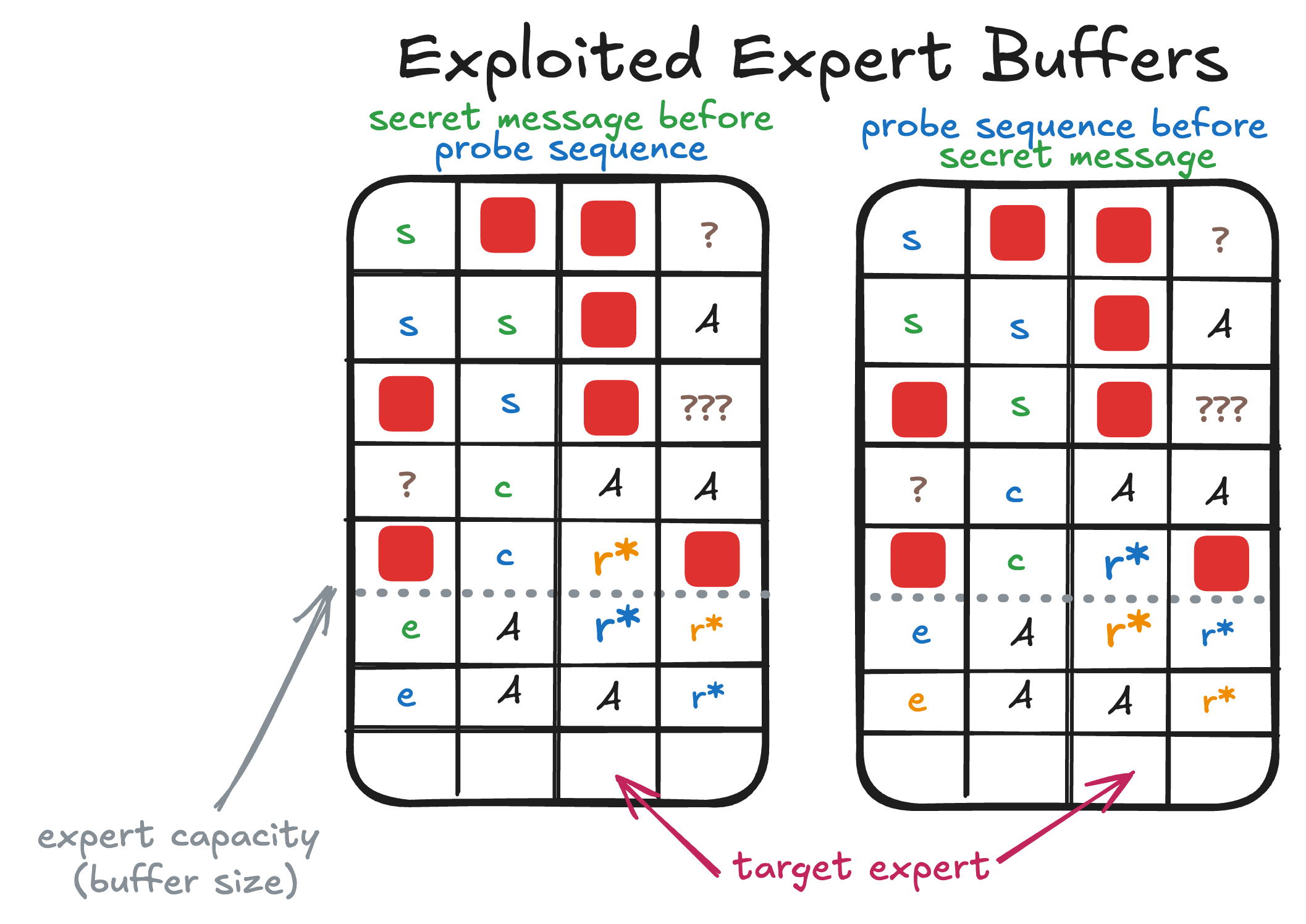}
  \end{center}
  \caption{The goal of the adversarial batch from \Cref{fig:adversarial_batch} is to carefully shape the expert buffers. This figure illustrates how the expert buffers looks like under successful exploitation that requires the knowledge of the target token and its position in the expert buffer. In this setting a change in the relative order of the \textbf{secret message} and the \textcolor{myblue}{\textbf{probe sequence}} will effect the routing decision and therefore the processing of the \textcolor{mygold}{target token} and the \textcolor{myblue}{guess token} as they are ties placed exactly at the edge of the expert buffer.}
\label{fig:exploited_expert_buffers}
\end{figure}

\subsection{Leakage Attack}
\label{ssec:attackoverview}
The adversary strategically crafts a batch of inputs, referred to as the \textit{adversarial batch}, to manipulate the expert routing within the MoE model. This manipulation forces the model to drop specific tokens from the victim's input, creating a detectable pattern that reveals information about the victim's prompt.

We detail the attack procedure in \Cref{highlevelalgo}. The steps involved are as follows:

\begin{enumerate}
    \item \textbf{Step 1: Guess the Next Token and its Position:} The attacker guesses the target token and its position in a chosen expert's buffer, assuming the prefix is known (initially empty).
    \item \textbf{Step 2: Construct the Adversarial Batch:} As illustrated in \Cref{fig:adversarial_batch} and using the primitives mentioned in \Cref{ssec:prims}, the attacker crafts an adversarial batch that:
    \begin{enumerate}
        \item Places \blockingsequence{}s to fill the expert buffer, leaving one spot for the guessed token; %
        \item Includes the \probeinput{} with the known prefix and guessed token, with the goal of triggering tie-handling; %
        \item Adds a \longsequence{} to extend the expect capacity. %
    \end{enumerate}

    \item \textbf{Step 3: Send Two Queries:} The attacker sends the adversarial batch twice, changing the order of the victim's message and the \probeinput{}.
    \item \textbf{Step 4: Map Observed Logits to Routing Paths:} The attacker uses a local model to find a mapping between observed logits to routing paths of the \probeinput{}, this is explained in depth in \Cref{sec:recoverroutingpath}.
    \item \textbf{Step 5: Verify the Guess:} A correct guess is indicated if the guessed target token is routed to the chosen expert in the first query (where the \probeinput{} comes first) but not in the second query (where the victim's message comes first).
\end{enumerate}

\subsection{Oracle Attack}
The Oracle Attack offers an efficient way to verify a guessed prompt when the attacker has complete knowledge of the candidate message. This eliminates the complexity of the general Leakage Attack, which requires iterative guessing and routing path recovery.

This simplification is achieved by:

\begin{itemize}
    \item \textbf{Targeting a single token}: The attack focuses solely on the last secret token, whose representation reflects the entire preceding sequence, thus a tie between target token and guess token will verify the complete candidate message at once.
    
    \item \textbf{Predicting token position}: With knowledge of the candidate message, the entire adversarial batch is known, and therefore the position of the target token within the expert buffer.
    
    \item \textbf{Bypassing routing path recovery}: With perfect knowledge of the candidate message, the attacker can deterministically predict token routing paths, avoiding the computationally expensive process of recovering them from the model's output.
\end{itemize}

By exploiting these observations, the Oracle Attack transforms the prompt verification problem into a direct assessment, removing the need for computationally expensive search of the Leakage Attack.

\subsection{Attack Complexity}
\label{ssec:attackcomplexity}
The Leakage Attack extracts the victim's prompt token-by-token. For each of the \seclen{} tokens in the victim's prompt, the attack iterates through the entire vocabulary (\vocabsize{}) and \seclen{} possible positions within the expert's buffer, resulting in \onlinecomplexity{} guesses or queries to the target model. Verifying each guess requires a computation of all $ 2 ^ {\text{\numlayers{}} \text{\numexperts{}}} $ routing paths (for a model with \numlayers{} layers and \numexperts{} experts), leading to \offlinecomplexity{} queries to a local copy of the model. The Oracle Attack, with its knowledge of the prompt, requires only two queries to the target model and its local copy.

\section{Evaluation}
\label{sec:evaluation}
\textbf{Setting} We evaluate our attack on the \textbf{first two transformer blocks} of \texttt{Mixtral-8x7B} \citep{jiang2024mixtralexperts}, using PyTorch 2.2.0+cu118. We set the model router to be Expert Choice Router as described by~\citet{zhou2024expertchoicerouter}. We restrict the vocabulary for guesses to lowercase letters and space, for a total of 27 tokens, and limit our extraction messages to the 1,000 most common words in English. We use a restricted vocab of only 9,218 out of 32,000 tokens for finding blockers, which we discuss in detail in~\Cref{sec:findingblockers}. We quantize the router weights to 5 digits to induce ties. We vary the length of padding sequences and enumerate all experts if needed to increase the success rate of our attack. We list evaluation parameters in \Cref{sec:notations}. Our evaluation focuses on a specific MoE model and a limited vocabulary. Further research is needed to assess the attack's effectiveness on different MoE architectures, larger vocabularies, and real-world deployment scenarios.

\textbf{\attackname{}} We find that it is possible to extract secret user data for all of the possible inputs we considered. We managed to fully extract 996 out of 1,000 secret messages and 4,833 out of 4,838 total secret tokens as shown in \Cref{tab:results}. We further explored how length of the user-message, the length of the padding sequence (which effects expert capacity), and the use of multiple experts affects the performance. %

\begin{table}[!htb]
\small
\begin{tabular}{l|ccccccccccc|r}
\toprule
\textbf{Secret message length} & 1 & 2 & 3 & 4 & 5 & 6 & 7 & 8 & 9 & 10 & 11 & \textbf{Total}\\
\textbf{Number of messages} & 2 & 25 & 125 & 329 & 236 & 148 & 78 & 40 & 10 & 6 & 1 & 1000 \\
\textbf{Total number of tokens} & 2 & 50 & 375 & 1316 & 1180 & 888 & 546 & 320 & 90 & 60 & 11 & 4838 \\
\textbf{Successfully recovered tokens} & 2 & 50 & 375 & 1315 & 1180 & 888 & 546 & 320 & 86 & 60 & 11 & 4833\\
\bottomrule
\end{tabular}
\caption{Attack performance across the 1,000 most common English words. These words were tokenized at the character level, resulting in a total of 4,838 individual tokens. By targeting all 8 experts in the first MoE layer and using 6 different padding sequence length our \attackname{} Attack successfully recovered 99.9\% (4,833) of them.}
\label{tab:results}
\end{table}

\begin{figure}[!htb]
    \centering
    \includegraphics[width=0.8\linewidth]{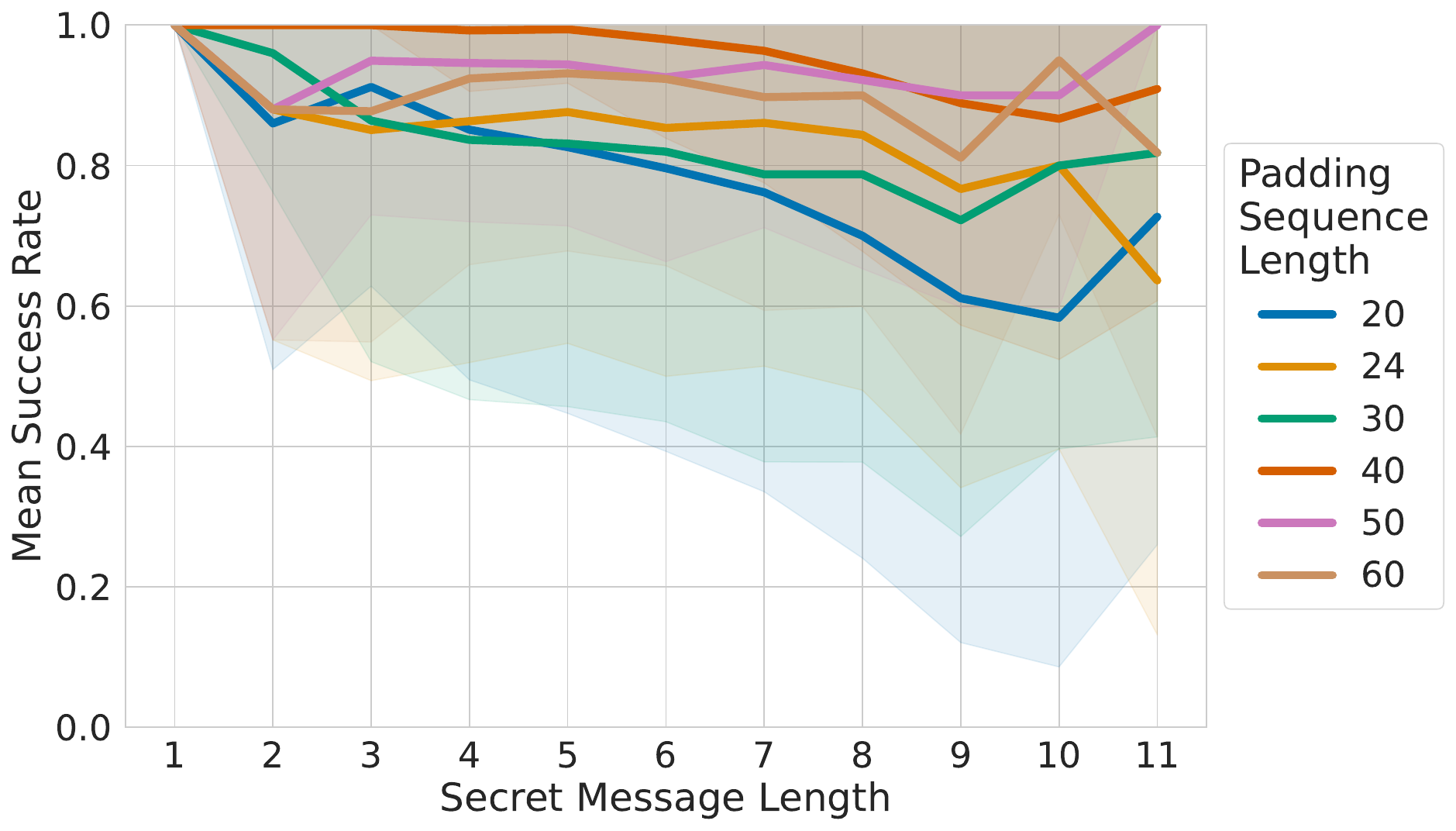}
    \caption{Attack performance of victim messages of different sizes per padding sequence length. The plot indicates there is a trade-off between padding sequence length and success rate, and that the attack becomes harder with the length of the secret message.} %
    \label{fig:performance_combined}
\end{figure}

\textbf{How expert capacity affects success rate?}
Expert capacity (buffer size) presents a trade-off: the bigger the buffer size is, the more likely the target token will be routed and not dropped, and potentially there is less interference between the adversarial batch and other victim (prefix/suffix) tokens. However, a longer expert capacity suggests more blocking is needed. A fixed batch size limits the number of blocking sequences, necessitating more blocking tokens per sequence. Finding such useful long blocking sequences is hard, thus the blocking sequences contain many non-blocking tokens which affects the routing of the victims tokens and in turn the reliability of the exploit. \Cref{fig:performance_combined} illustrates this trade-off, as the optimal padding sequence length that is used to control the expert capacity is 40 in our evaluation. With that we are able to leak with 100\% all messages of length 5 which are the vast majority of the messages in our dataset. There is a small positive correlation between message length and adequate padding sequence length, this is probably crucial for ensuring all tokens are not dropped by default.

\begin{figure}[!htb]
    \centering
    \includegraphics[width=0.8\linewidth]{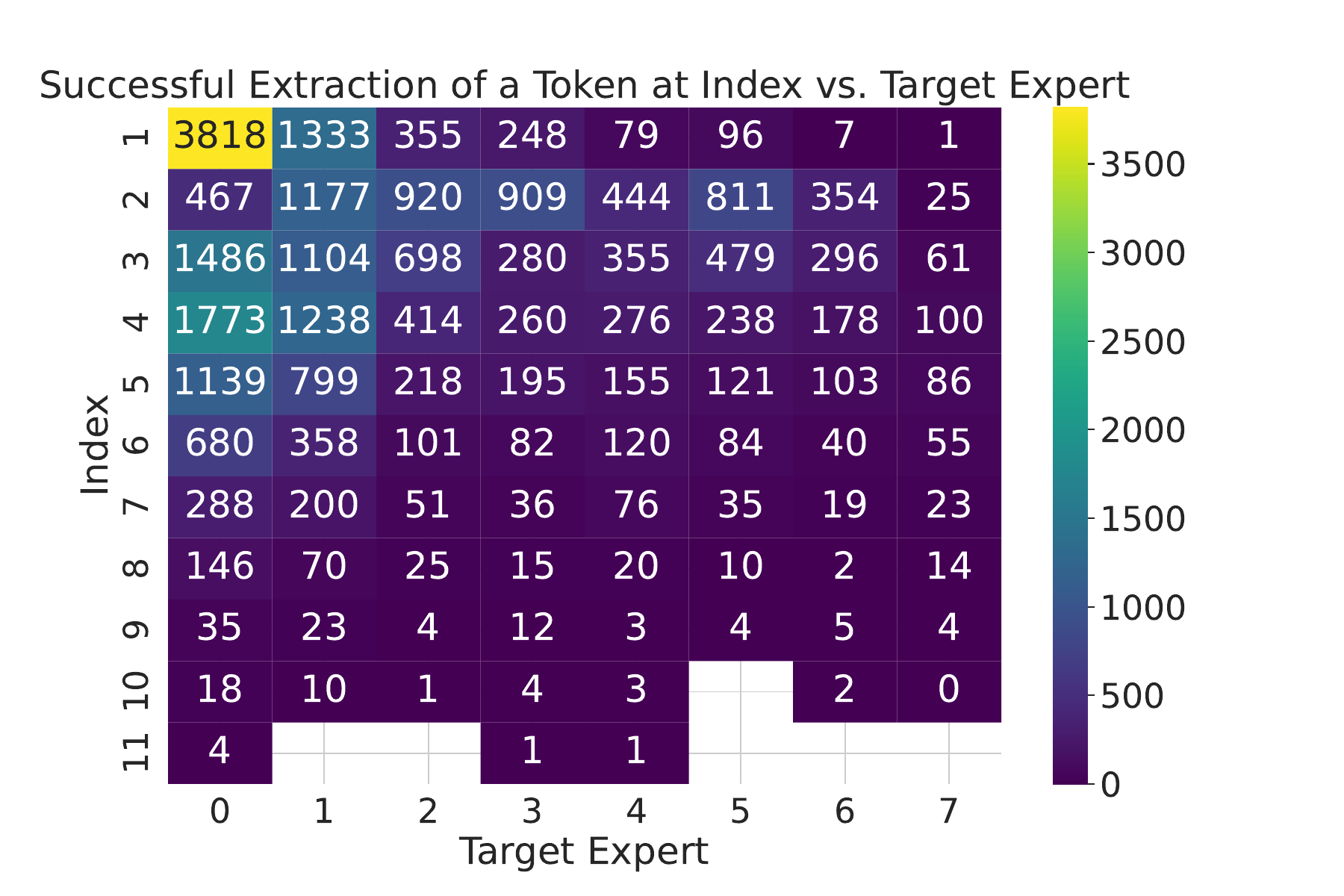}
    \caption{Heatmap showing the correlation between the expert and the index of the input token where the attack succeeds. Here, the attack progresses to the next token when any expert is successfully exploited to leak the token of the victim. There is a diminishing utility in using all experts, but they were all necessary to achieve a success rate > 99\%.}
    \label{fig:heatmap_index_vs_targetexpert}
\end{figure}

\textbf{Which experts leak the victim tokens?} \Cref{fig:heatmap_index_vs_targetexpert} shows the expert and the index where the attack succeeds. Note that \attackname{} moves to the next token when a successful attack is found on any of the experts, meaning that the plot shows the index and the target expert that tend to work first for extraction. %

\begin{figure}[!htb]
    \centering
    \includegraphics[width=0.8\linewidth]{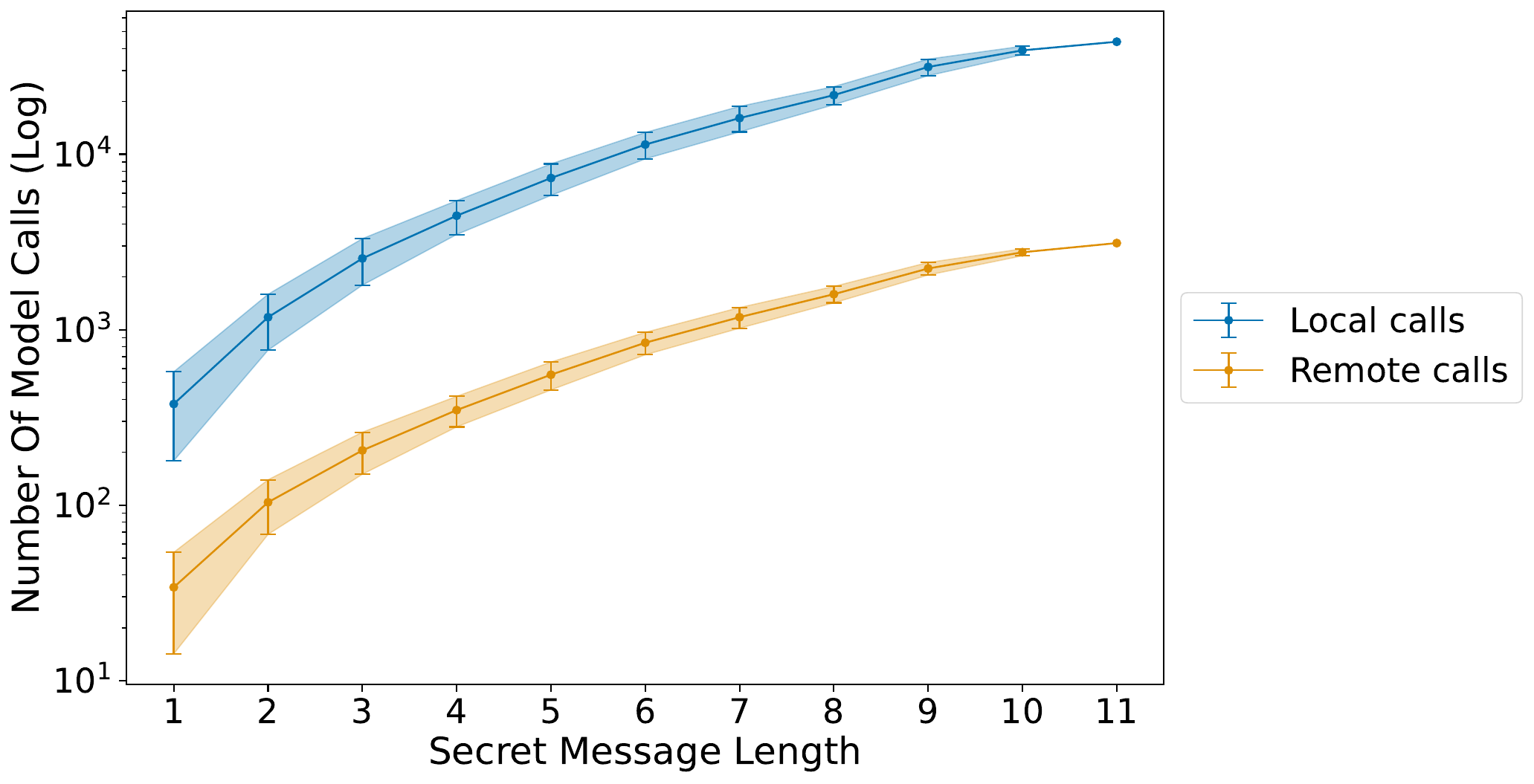}
    \caption{Number of model calls required to leak the victim's secret message. The majority of calls are performed locally by the adversary on the local copy of the model, with only a fraction of remote calls performed by the target model required to execute the attack. We find that the attack requires up to a 100 queries on average to the target model per token. The trend of the two graphs is a function of \onlinecomplexity{} which is shared between the local and remote query complexity. The fixed gap is a function of \numlayers{} the number of layers and \numexperts{} experts, further explained in \Cref{ssec:routingpath_necessaity}.  }
    \label{fig:numberofcalls}
\end{figure}
\FloatBarrier{}

\section{Discussion}
\label{sec:discussion}

\textbf{Methodological improvements}: In this paper we show that it is possible to exploit \ecr{} to extract private victim data placed in the same batch as an adversary's data.
\attackname{} currently requires 2 queries per guess for verification, and $2^{\text{\numlayers\numexperts}}$ per-token queries for general extraction. This makes it infeasible at present to use our attack against real world systems.
However, we believe that performance of the attack could be significantly improved. 
First, we hypothesize that refining the buffer shaping process could enable the selection of blockers that prevent inter-batch token interference. We discuss this further in~\Cref{sec:findingblockers}. 
Second, we suspect that an alternative approach exists to determine the processed token without exhaustively constructing all possible  $2^{\text{\numlayers\numexperts}}$ expert combinations, potentially by learning a mapping between outputs and routing paths. 
We discuss this further in~\Cref{sec:recoverroutingpath}. 
Third, targeting the final MoE layer instead of the first may eliminate the need for routing path tracking altogether. 
Fourth, the current attack requires precise matching in its exploitation for tie-handling. 
We hypothesize that relative placement of tokens can similarly be used to signal what victim token is used. 
Fifth, a single query oracle attack might be possible by sending multiple probe inputs. Sixth, we believe might be possible to reduce \onlinecomplexity{} using binary-search over the target token position, and by being able to leak information from non-equal priorities, or by simply using an LLM to propose guesses for the next token instead of naively trying them in order.
Finally, we believe that a black-box variant of this attack is feasible -- a local clone of the model at present is only used to find blocking sequences and for inverting token routing paths. 
We hypothesize that both of the tasks could inefficiently be deduced from black-box access.

\textbf{Optimisation--Security Trade-Off}: Within the domain of computer security, it is well-established that prioritizing performance optimization often inadvertently introduces vulnerabilities to side-channel attacks~\citep{anderson2010security}. 
In the case of MoE models, the \ecr{} strategy, which optimizes for efficiency, inadvertently creates a side channel that allows the attacker to exploit the model. 
Our work highlights the importance of rigorous adversarial testing of any optimization introduced into machine learning pipelines to safeguard user privacy. 
While we focus on a specific routing strategy, we anticipate that similar vulnerabilities may exist in other strategies that violate implicit batch independence.

\textbf{Defenses}: Having established a general vulnerability of MoE-based models with \ecr{}, we now shift our focus to potential defense strategies. 
A crucial first step in mitigating these vulnerabilities is to preserve in-batch data independence, particularly across different users. 
This ensures that adversaries cannot exploit the routing strategy. 
Second, the current attack design requires precise shaping of the expert buffer; therefore, introducing any form of stochasticity into the system can effectively disrupt the attacker's ability to exploit this vulnerability.
This could involve incorporating randomness into various aspects of the model, such as the expert capacity factor, the batch order, the input itself, or the routing strategy.

\section{Related Work}

\subsection{\moe{}}
The concept of \moe{} (MoE) was first introduced by \citet{jacobs1991adaptivemix} and \cite{jordan1994hierarchical}, but has more recently become a popular tool for efficient inference on Transformer-based LLMs~\citep{shazeer2017outrageously,zoph2022stmoedesigningstabletransferable,renggli2022learningmergetokensvision,jiang2024mixtralexperts}, primarily because it allows the model to activate only a small fraction of its total parameters for any given input.
An MoE layer in a large language model (LLM) consists of \numexperts{} expert modules and a gating function, which routes a token to an expert (or subset of the \numexperts{} experts).
Since only a subset of experts is activated for an input token, the number of parameters activated is significantly smaller than the overall number of parameters in the LLM, which translates to fewer floating-point operations and faster inference.
This in turn allows one to build extremely large networks without a corresponding increase in inference costs. Some of the best performing modern LLMs utilize MoE architectures e.g. Gemini-1.5 \citep{geminiteam2024geminifamilyhighlycapable}
, Mixtral~\citep{jiang2024mixtralexperts}, and Grok-1~\citep{grok1}.

\subsection{Violations of User Privacy}
Although previous work has investigated how user privacy can be compromised in LLMs~\citep{debenedetti2023privacy,shen2024prompt}, none have thus far investigated vulnerability of user privacy due to the underlying model architecture, specifically how input representations can be influenced by other data within the same batch.
\cite{hayes2024buffer} demonstrated previously that batch composition can be used by adversaries to exploit MoE routing and to launch Denial-of-Service (DoS) attacks; we instead exploit it to leak private user supplied prompts. 
Note that \ecr{} considered in this work is one of many different routing strategies that breaks the implicit batch independence; we hypothesise that other routing strategies may be similarly vulnerable.

\section{Conclusion}

In classical dense LLMs, it is essentially impossible for one user's data to impact another user's output.
But MoE models introduce a side-channel: one user's queries can impact a different user's outputs.
The magnitude of this leak is minuscule and challenging to detect.
But by carefully crafting adversarial input batches, we show how to manipulate the expert buffers within the MoE model with \ecr{}, leading to the full disclosure of a victim’s prompt included in the same batch. At present, \attackname{} is only possible when \ecr{} is used, yet we hypothesise that other routing strategies can be similarly vulnerable.
While the current threat model assumes unrealistic attacker capabilities, we believe that future research can extend the practicality of these attacks. 

More broadly, attacks such as this highlight the importance of system-level security analysis at
all stages of model deployment, starting as the design of the architecture, and extending towards
as late as the actual deployment of the model and how different user queries are batched together.
Studying any one component in isolation may give the appearance of safety, but only when the system
as a whole is analyzed is it possible to understand vulnerabilities such as this.
We hope that future work will perform other analysis of this type on future advances.

\section{Reproducibility Statement}
To ensure reproducibility, we provide a comprehensive outline of our attack methodology in \Cref{sec:methodology}. We include all of the details about the attack and provide a detailed algorithmic description in \Cref{ssec:prims}. Our evaluation in \Cref{sec:evaluation} relies on the base model that is openly available (Mixtral-8x7B). A number of supplementary figures in the appendix illustrate all of the details required to replicate the work. We list the hyperparameters and the code for the router in the Appendix.

\bibliographystyle{abbrvnat}
\nobibliography*
\bibliography{template_refs}

\appendix
\FloatBarrier

\section{Notations | List of parameters and their values}
\label{sec:notations}
\FloatBarrier
\setlength\cellspacetoplimit{4pt}  %
\setlength\cellspacebottomlimit{0pt}  %

\begin{table}[!htb]
\centering
\adjustbox{width=\linewidth}{
\begin{tabular}{Sl Sl Sl >{\raggedright\arraybackslash}p{6cm}}
\toprule
\footnotesize Notation & \footnotesize Name & \footnotesize Value(s) & \footnotesize Description \\
\midrule
\seclen{} & Secret message length & 1 - 11 & The number of tokens in the victim's secret message. \\
\numlayers{} & Model depth & 2 &  The number of layers in the model, restricted to the first two layers of Mixtral in our evaluation. \\
 & Quantisation param & 5 &  The number of digits used to round attention outputs to ensure ties. \\
 
\vocabsize{} & Guess vocabulary size & $|\mathcal{V}| = 27$ & The number of tokens the attacker considers when making guesses. In our evaluation, this is restricted to lowercase letters and space. \\

\numexperts{} & Number of experts & 8 & The number of experts per layer. This is the default Mixtral setting. \\

\batchsize{} & Batch size & 32 & The number of sequences in a given batch, including the victim's secret message. This is a typical batch size value. \\
\capacityfactor{} & Capacity factor & 1.0 & A parameter in \ecr{} that determines the portion of tokens each expert will process relative to the average. \\
\padlen{} & Padding sequence length & \{20, 24, 30, 40, 50, 60\} & The length of the padding sequence used to control max sequence length and with that extend the expert buffer size. \\
\seqlen{} & Max sequence length & $max(\text{\seclen{}},~\text{\padlen{}})$ & The maximum length of any sequence in the adversarial batch, determined by the longer of the victim's message (\seclen{}) and the padding sequence (\padlen{}). \\
\expertcapacity{} & Expert capacity & \{80, 96, 120, 160, 200, 240\} & The number of tokens each expert processes. This is can be controlled by the attacker using a padding sequence. \\
\bseqT{} & Blocker sequence threshold & 0.85 & Minimum priority for a token to be considered a blocker. \\
  & Blocker vocabulary size & 9,218 & Subset of Mixtral's vocabulary (32,000 tokens) used for blocking, ensuring no token merging. \\
$\beta$ & Max paths Hamming-distance & 4 & A heuristic used to reduce routing paths we enumerate. An initial routing path is estimated and only paths that differ by up to $\beta$ bits are recorded. \\\\
\bottomrule
\end{tabular}}
\caption{Table of Notations}
\label{table:notations}
\end{table}
\FloatBarrier

\section{\attackname{} Algorithm}\label{highlevelalgo}
\FloatBarrier
\begin{algorithm}[!htb]

\DontPrintSemicolon %
\KwIn{Tokens Vocabulary \vocab{}, Number of experts \numexperts{}, Number of layers \numlayers{}, Capacity factor \capacityfactor{}, Victim's message length \seclen{}, batch size \batchsize{}}
\KwOut{Victim's message}
$ expert \gets 0$ \tcp{fix expert to target, can also be enumerated}
$ prefix \gets \emptyset$ \tcp{this is the prefix known to the attacker}
$ params \gets (\text{\vocab{}},
                \text{\numexperts{}},
                \text{\numlayers{}},
                \text{\capacityfactor{}},
                \text{\seclen{}},
                \text{\batchsize{}})$ \;

\BlankLine
\tcp{extract token-by-token}
\For{$token\_index \gets 1$ \KwTo $M$}{ \BlankLine \tcp{guess next token}
    \For{guess in \vocab{}}{
        $ probe\_seq \gets prefix + guess $ \;
        $ min\_position \gets get\_minimal\_position(probe\_sequence,~params) $ \;
    
            \BlankLine \tcp{iterate over all possible positions for the token}
            \For{$position \gets min\_position$ \KwTo $min\_position + M$}{

                $\phantom{}{\advbatch{}} \gets construct\_adv\_batch($\offlinemodel$, probe\_seq, position, expert, params)$ \;
                $out1 \gets \onlinemodel(\advbatch, \usersec)$ \;
                $out2 \gets \onlinemodel(\usersec, \advbatch)$ \;
                
                \BlankLine \tcp{collect all possible logits for different droppings}
                $routing\_paths \gets logits\_to\_routing\_paths(\offlinemodel, \advbatch, probe\_seq)$ \;
                
                \BlankLine \tcp{not dropped = 1, dropped = 0 <=> guess is correct}
                \If{$routing\_paths[out1][expert] >  routing\_paths[out2][expert] $}{
                    $prefix \gets prefix + guess$ \;
                    
                    \textbf{break} \tcp{break out of positions}
                
            }
        }
    }
}
\caption{{High-level \attackname{} algorithm}}
\end{algorithm}
\FloatBarrier

\section{Exploiting Tie-Handling in \texttt{Topk} Implementations}\label{sec:topkdeterminsm}
Our attack relies on the consistent and stable tie-handling behavior of the \topk{} implementation in PyTorch 2.2.0+cu118 with a CUDA environment. However, this behavior is not guaranteed on CPUs, where  \topk{} can produce inconsistent results for duplicate elements \citep{pytorchissue2,pytorchissue3,pytorchissue1}. As demonstrated in the code listing below, \topk{} reliably returns ordered indices for CUDA tensors (green highlights) but fails to do so for CPU tensors (red highlights). This necessitates an alternative approach for exploiting ties on CPUs. 

\FloatBarrier
\begin{minted}[fontsize=\small]{Python}
import torch
for device in ['cuda', 'cpu']:
  for size in [32, 33]:
    for is_sorted in [True, False]:
      print(size, is_sorted, device)
      print(torch.topk(torch.Tensor([1] * size).to(device), 
            k = size, sorted = is_sorted).indices)
\end{minted}
\FloatBarrier

\definecolor{darkgreen}{HTML}{00bb55}
\definecolor{darkred}{HTML}{e90000}

\def\speciallstcolor{\begingroup\color{darkgreen}}
\def\endspeciallstcolor{\endgroup}

\texttt{Output:}
\begin{lstlisting}[linewidth=\linewidth, basicstyle=\scriptsize\ttfamily,escapechar=$]
32 True cuda
$\colorbox{darkred}{\color{white}{tensor([31, 30, 28, 29, 25, 24, 26, 27, 19, 18, 16,}}\\
\colorbox{darkred}{\color{white}{17, 21, 20, 22, 23, 
7,  6, 4,  5,  1,  0,  2,  3, 11, 10,  8,  9, 13, 12, 14, 15], device='cuda:0')}}$
32 False cuda
$\colorbox{darkgreen}{\color{white}{tensor([ 0,  1,  2,  3,  4,  5,  6,  7,  8,  9, 10, 11, 12, 13, 14, 15, 16,}}\\\colorbox{darkgreen}{\color{white}{
17, 18, 19, 20, 21, 22, 23, 24, 25, 26, 27, 28, 29, 30, 31], device='cuda:0')}}$
33 True cuda
$\colorbox{darkgreen}{\color{white}{tensor([ 0,  1,  2,  3,  4,  5,  6,  7,  8,  9, 10, 11, 12, 13, 14, 15, 16,}}\\\colorbox{darkgreen}{\color{white}{
17, 18, 19, 20, 21, 22, 23, 24, 25, 26, 27, 28, 29, 30, 31, 32], device='cuda:0')}}$
33 False cuda
$\colorbox{darkgreen}{\color{white}{tensor([ 0,  1,  2,  3,  4,  5,  6,  7,  8,  9, 10, 11, 12, 13, 14, 15, 16,}}\\\colorbox{darkgreen}{\color{white}{ 
17, 18, 19, 20, 21, 22, 23, 24, 25, 26, 27, 28, 29, 30, 31, 32], device='cuda:0')}}$
32 True cpu
$\colorbox{darkred}{\color{white}{tensor([17,  0,  9, 10, 13, 14, 15, 12,  7,  6,  5,  4,  3,  2,  1,  8,}}\\
\colorbox{darkred}{\color{white}{16, 18, 19, 20, 21, 22, 23, 24, 25, 26, 27, 28, 29, 30, 31, 11])}}$
32 False cpu
$\colorbox{darkred}{\color{white}{tensor([16, 31, 30, 29, 28, 27, 26, 25, 24, 23, 22, 21, 20, 19, 18, 17, }}\\\colorbox{darkred}{\color{white}{
8,  1, 2,  3,  4,  5,  6,  7, 12, 15, 14, 13, 10,  9,  0, 11])}}$
33 True cpu
$\colorbox{darkred}{\color{white}{tensor([17,  0,  9, 10, 13, 14, 15, 12,  7,  6,  5,  4,  3,  2,  1,  8,}}\\
\colorbox{darkred}{\color{white}{16, 18, 19, 20, 21, 22, 23, 24, 25, 26, 27, 28, 29, 30, 31, 32, 11])}}$
33 False cpu
$\colorbox{darkred}{\color{white}{tensor([16, 32, 31, 30, 29, 28, 27, 26, 25, 24, 23, 22, 21, 20, 19,}}\\
\colorbox{darkred}{\color{white}{18, 17,  8, 1,  2,  3,  4,  5,  6,  7, 12, 15, 14, 13, 10,  9,  0, 11])}}$
\end{lstlisting}

Notice also that \texttt{torch.}\topk{} outputs for duplicates are predictable (sorted) without setting sort=True for CUDA tensors of $size > 32$. This too is an implementation detail crucial for our attack, torch.\topk{} is based  on sorting the buffer and returning the first k elements of it, where different sorting algorithms are used for different buffer sizes. For buffers with $ size <= 32$ the implementation uses the unstable \textit{bitonic sort}, and for larger sizes it uses stable versions of other sorting algorithms such as: merge sort and radix sort. This can be inferred from PyTorch implementation~\citep{torchimpl}, in /aten/src/ATen/native/cuda/Sort.cu:

\begin{minted}[fontsize=\small]{C++}
void sortKeyValueInplace(
    const TensorBase& key,
    const TensorBase& value,
    int64_t dim,
    bool descending,
    bool stable) {
  const auto sort_size = key.size(dim);
  if (sort_size <= 1) {
    return; // Already sorted
  } else if (!stable && sort_size <= 32) {
    // NOTE: Bitonic sort is unstable
    sortCommon(SmallBitonicSort{}, key, value, dim, descending);
#if HAS_WARP_MERGE_SORT()
  } else if (sort_size <= 128) {
    sortCommon(WarpMergeSort<128>{}, key, value, dim, descending);
#endif
  } else {
    sortCommon(MediumRadixSort{}, key, value, dim, descending);
  }
}
\end{minted}
To ensure predictable tie-handling in \texttt{torch.topk}, we extend the expert capacity (the buffer size for \topk) to be greater than 32 with a \longsequence{}.

\section{Find Blocking Sequences}
\label{sec:findingblockers}

To construct adversarial batches efficiently, the attacker must generate \blockingsequence{}s for each expert. These sequences consist of high-priority tokens that, when included in the batch, fill the expert's buffer up to a desired threshold. This process involves the following steps:

\begin{enumerate}
    \item \textbf{Vocabulary Restriction:}  Use a restricted vocabulary \bvocab{} of prefix-free tokens. This prevents unintended token merging when combining blocking sequences.
    \item \textbf{Priority Threshold:} Set a priority threshold, denoted as  \bseqT, to 0.85. This value determines the minimum priority for a token to be considered a blocker.
    \item \textbf{Blocker Limit:}  Define $nb$ as the maximum number of blocking tokens allowed in a single blocking sequence. This is calculated as  $(\expertcapacity - 1) / (\batchsize - 3)$,  ensuring that the total number of blockers in the batch does not exceed the expert's capacity.
    \item \textbf{Sequence Generation:} Generate a blocking sequence with length $bsl$ (where $bsl \leq \text{\padlen{}}$) containing $nb$ tokens, each with a priority $p_{e_i}$ less than \bseqT{} for the target expert $e_i$.
\end{enumerate}

The algorithm for generating a blocking sequence is as follows:

\begin{enumerate}
    \item Initialize an empty \blockingsequence{} with the beginning-of-sequence token ($<$bos$>$).
    \item Randomly generate a candidate chunk of length $bsl / nb$.
    \item If the chunk contains at least one token with priority $p_{e_i} \geq \text{\bseqT}$, append the chunk to the \blockingsequence.
    \item Trim any unnecessary tokens from the end of the \blockingsequence.
    \item Repeat steps 2-4 until $nb$ chunks have been appended.
\end{enumerate}

\section{Recovering A Token Routing Path}\label{sec:recoverroutingpath}
We define a token's \textit{routing path} for \ecr{} as a binary matrix $R$ of shape $(\text{\numexperts} \times \text{\numlayers})$, where $R_{ij} = 1$ if the token is routed to expert $e_i$ in layer $j$, and 0 otherwise. This matrix encodes the specific sequence of experts that a token is assigned to as it passes through the model's layers. 

During \attackname{} we query the model with two \textit{adversarial batches} corresponding to a guess of the target token and its position in the expert buffer. The two queries only differ in the position of the \probeinput{}. We expect to observe changes in the model's output for the \probeinput{} in the two queries if the guess is correct. However, these output variations could stem from several factors:

\begin{itemize}
    \item \textbf{Numerical instability}: Floating-point errors inherent in computations.
    \item \textbf{Suffix $\rightarrow$ prefix dropout}: As attackers we only have partial knowledge of the state of the expert because of the suffix, or victim's message continuation which is unknown to us. This suffix could fill other experts and cause the to drop prefix tokens.
    \item \textbf{Double dropout}: Simultaneous dropping of identical tokens at the boundaries of different expert buffers. In \probeinput{} we repeat both the target token and all the prefix tokens, therefore they would all have identical representations.
\end{itemize}

To avoid misinterpreting these variations as successful attacks (false positives), we need to ensure that any observed changes are specifically due to the intended conditional dropout into the target expert.
Our proposed approach is robust but computationally expensive, limiting its scalability to models with more than two MoE layers. This computational cost arises from the need to meticulously track token routing paths.

Here is how it works:
\begin{enumerate}
    \item \textbf{Establish prefix path}: We first estimate the routing path of prefix tokens and cache their attention activations using KV-caching. This is done by processing the adversarial batch without the target token.
    \item \textbf{Explore Potential Paths}: For each possible expert allocation of the target token ($2^{\text{\numexperts} \times \text{\numlayers}}$ combinations), we query a local model with the target token and the cached activations. This local model allows us to selectively disable experts based on a given \emph{routing path}.
    \item \textbf{Map Outputs to Paths}:  We create a lookup table that maps each model output to its corresponding routing path. This "routing path table" helps us trace how tokens are assigned to experts.
    \item \textbf{Analyze Adversarial Outputs}: When we query the target model with the adversarial batch, we compare the resulting outputs to our routing path table. This allows us to reconstruct the actual routing path taken by the target token. Ideally, we expect to see a distinct pattern where the \probeinput{} in the first adversarial batch is not dropped.
\end{enumerate}

To account for potential floating-point discrepancies, we use approximate nearest neighbor search in the routing path table based on an $L_0$ distance metric, rather than relying on exact matches.

\subsection{Are Token Routing Path Maps Necessary?}\label{ssec:routingpath_necessaity}
It might seem that explicitly computing token routing paths is an unnecessary step. However, this is crucial for our attack, especially in models with multiple MoE layers.

Consider a single-layer MoE model (Embedding, Attention, MoE, Unembedding), the attention mechanism doesn't modify the representation of a token after it's been processed by the MoE layer. This means that any token dropout in the MoE layer directly impacts the model's final output (logits), without any further interaction or information flow between tokens through attention.

In contrast to a single-layer model, any deeper model, for example a two-layer MoE model (Embedding, Attention, MoE, Attention, MoE, Unembedding) introduces complexities. The second attention layer is influenced by token dropouts occurring in the \textit{first} MoE layer. This means a dropped token will have a modified representation due to the dropout itself, the subsequent attention mechanism, and potentially the dropouts in the following MoE layers.
Inevitably, we need to track the target token's representation throughout the entire network to accurately link the model's output back to the tie-breaking behavior in the first MoE layer. 

In our evaluation, in order to extract thousands of tokens we restricted the set of routing paths using a heuristic. We estimated the target token routing path and only mapped nearby paths (ones that differ by at most $\beta = 4$ bits from the estimated path). Additionally, when the returned logits of the two queries were close enough we skipped the routing path enumeration all together, assuming the guess was incorrect. That was sufficient in our limited setting, but, it does not affect the scalability of the attack with respect to the number of layers in a general setting.

\section{\ecr{} implementation with Mixtral}
The following code provides an implementation of: (a) Slicing a Mixtral model to two layers, (b) \ecr{}, (c) Loading and Hooking models to use \ecr{}. 

\begin{minted}
[
breaklines,
fontsize=\tiny
]
{python}
def slice_model(model_name = 'mistralai/Mixtral-8x7B-Instruct-v0.1', num_layers = 2):

    # load full model
    full_model = transformers.AutoModelForCausalLM.from_pretrained(model_name)
    cfg = copy.deepcopy(full_model.config)
    
    # slice
    cfg.num_hidden_layers = num_layers
    sliced_model = transformers.AutoModelForCausalLM.from_pretrained(model_name, config=cfg)
    
    # store
    sliced_model.save_pretrained(f'./sliced_{model_name.replace("/", "_")}_{num_layers}_layer{"s" if num_layers > 1 else ""}')
    
# Same class but pass attention_mask to MoE layer so it can deprioritize <pad> tokens.
class ModifiedLayer(transformers.models.mixtral.modeling_mixtral.MixtralDecoderLayer):
    def forward(
        self,
        hidden_states: torch.Tensor,
        attention_mask: Optional[torch.Tensor] = None,
        position_ids: Optional[torch.LongTensor] = None,
        past_key_value: Optional[Tuple[torch.Tensor]] = None,
        output_attentions: Optional[bool] = False,
        output_router_logits: Optional[bool] = False,
        use_cache: Optional[bool] = False,
    ) -> Tuple[torch.FloatTensor, Optional[Tuple[torch.FloatTensor, torch.FloatTensor]]]:
        """
        Args:
            hidden_states (`torch.FloatTensor`): input to the layer of shape `(batch, seq_len, embed_dim)`
            attention_mask (`torch.FloatTensor`, *optional*): attention mask of size
                `(batch, sequence_length)` where padding elements are indicated by 0.
            past_key_value (`Tuple(torch.FloatTensor)`, *optional*): cached past key and value projection states
            output_attentions (`bool`, *optional*):
                Whether or not to return the attentions tensors of all attention layers. See `attentions` under
                returned tensors for more detail.
            output_router_logits (`bool`, *optional*):
                Whether or not to return the logits of all the routers. They are useful for computing the router loss, and
                should not be returned during inference.
            use_cache (`bool`, *optional*):
                If set to `True`, `past_key_values` key value states are returned and can be used to speed up decoding
                (see `past_key_values`).
        """

        residual = hidden_states

        hidden_states = self.input_layernorm(hidden_states)

        # Self Attention
        hidden_states, self_attn_weights, present_key_value = self.self_attn(
            hidden_states=hidden_states,
            attention_mask=attention_mask,
            position_ids=position_ids,
            past_key_value=past_key_value,
            output_attentions=output_attentions,
            use_cache=use_cache,
        )
        hidden_states = residual + hidden_states

        # Fully Connected
        residual = hidden_states
        hidden_states = self.post_attention_layernorm(hidden_states)
        hidden_states, router_logits = self.block_sparse_moe(hidden_states, attention_mask) # the only change is to pass attention_mask to MoE block
        hidden_states = residual + hidden_states

        outputs = (hidden_states,)

        if output_attentions:
            outputs += (self_attn_weights,)

        if use_cache:
            outputs += (present_key_value,)

        if output_router_logits:
            outputs += (router_logits,)
            
        return outputs

class ExpertRoutingStrategy(transformers.models.mixtral.modeling_mixtral.MixtralSparseMoeBlock):
    ''' Based on: https://arxiv.org/abs/2202.09368, Section 3.2
        jax implementation is available here: https://github.com/google/flaxformer/blob/main/flaxformer/architectures/moe/routing.py#L647-L717
    
        n = total_number_of_tokens = batch_size * seq_length
        c = capacity factor # ~on average how many experts are utilized by a token
        e = number of experts
        k = n*c/e
        d = model_hidden_dim
        
        X = hidden_states => <n, d>
        outputs: I, G, P

        I[i,j] = the j_th selected token of the i_th expert => <e, k>
        G = the weight of expert for the selected token <e, k>
        P = one hot encoding of I => <e, k, n>
    '''
    capacity_factor = CAPACITY_FACTOR
    
    def forward(self, hidden_states: torch.Tensor, attention_mask: Optional[torch.Tensor]) -> torch.Tensor:
        batch_size, sequence_length, hidden_dim = hidden_states.shape
    
        if self.training and self.jitter_noise > 0:
            hidden_states *= torch.empty_like(hidden_states).uniform_(1.0 - self.jitter_noise, 1.0 + self.jitter_noise)

        hidden_states = hidden_states.view(-1, hidden_dim) # X <n, d>
        router_logits = self.gate(hidden_states)

        routing_weights = F.softmax(router_logits, dim=-1, dtype=torch.float) # S <n, e>
        
        # simplifying assumption
        if ROUND_ASSIST:
            routing_weights = routing_weights.round(decimals=ROUND_BY)

        # deprioritizing padding tokens
        if attention_mask is not None:
            padding_mask = attention_mask[:,0,-1].exp()
            routing_weights *= padding_mask.unsqueeze(-1).reshape(-1, 1)
        
        tokens_per_expert_on_avg = batch_size * sequence_length / self.num_experts 
        expert_capacity = int(tokens_per_expert_on_avg * self.capacity_factor)
        

        # store aside stuff for later probing
        self.routing_weights = routing_weights.clone()
        routing_weights, selected_tokens = torch.topk(routing_weights, k=expert_capacity, dim=0) # G, I
        self.routing_weights_topk = routing_weights.clone()
        self.selected_tokens = selected_tokens.clone()
    
        routing_weights = routing_weights.to(hidden_states.dtype)

        final_hidden_states = torch.zeros(
           (batch_size * sequence_length, hidden_dim), dtype=hidden_states.dtype, device=hidden_states.device
        )

        for expert_idx in range(self.num_experts):
            expert_layer = self.experts[expert_idx]
            token_idx = selected_tokens[:,expert_idx]

            current_state = hidden_states[None, token_idx].reshape(-1, hidden_dim)
            current_hidden_states = expert_layer(current_state) * routing_weights[:, expert_idx, None]

            final_hidden_states.index_add_(0, token_idx, current_hidden_states.to(hidden_states.dtype))
        final_hidden_states = final_hidden_states.reshape(batch_size, sequence_length, hidden_dim)
        return final_hidden_states, router_logits

def load_models(path, tokenizer_path, load_all_models = True):
   
    cfg = transformers.AutoConfig.from_pretrained(path)
    tokenizer = transformers.AutoTokenizer.from_pretrained(tokenizer_path)
    
    offline_model = offline_bitmap_model = online_model = None
    
    offline_model = transformers.AutoModelForCausalLM.from_pretrained(path)
    offline_model.config._name_or_path = None # avoid loading weights again

    if load_all_models:
        offline_bitmap_model = transformers.AutoModelForCausalLM.from_pretrained(path)
        offline_bitmap_model.config._name_or_path = None

        print_verbose("Loading online model", required_level=VerbosityLevel.DEBUG)
        online_model = transformers.AutoModelForCausalLM.from_pretrained(path)
        online_model.config._name_or_path = None

    models_and_hooks = [(offline_model, ExpertRoutingStrategy),
                        (offline_bitmap_model, ExpertRoutingStrategyWithBitmap),
                        (online_model, ExpertRoutingStrategy)]
    
    if not load_all_models:
        models_and_hooks = models_and_hooks[:1]
    
    for model, routing_strategy_cls in models_and_hooks:
        for i in trange(cfg.num_hidden_layers):
            modified_moe_layer = routing_strategy_cls(config=model.config)
            modified_layer = ModifiedLayer(config=model.config, layer_idx=i)
        
            # use original weights
            modified_layer.load_state_dict(model.model.layers[i].state_dict())
            modified_moe_layer.load_state_dict(model.model.layers[i].block_sparse_moe.state_dict()) 
        
            model.model.layers[i] = modified_layer
            model.model.layers[i].block_sparse_moe = modified_moe_layer

        # set hook to count forward calls
        model.forward = count_calls(model.forward)
        
    return cfg, tokenizer, offline_model, offline_bitmap_model, online_model
\end{minted}

\end{document}